\documentclass[twocolumn,twoside,slac_two]{revtex4}
\usepackage{graphicx}
\usepackage{fancyhdr}
\begin{document}

\title{Highly Accurate Analytic Presentation of Solution of the Schr\"{o}dinger
Equation}

%

\author{E. Z. Liverts}
\author{E.~G.~Drukarev}
\affiliation{Racah Institute of Physics, The Hebrew University, Jerusalem 91904, Israel}
\author{R.~Krivec}
\affiliation{J. Stefan Institute Jamova 39, 1000 Ljubljana, Slovenia}
\author{V.~B.~Mandelzweig}
\affiliation{Racah Institute of Physics, The Hebrew University, Jerusalem 91904, Israel}

\begin{abstract}
High-precision approximate analytic expressions for 
energies and 
wave functions 
are found 
for arbitrary physical potentials. The Schr\"{o}dinger equation is 
cast into 
nonlinear Riccati equation, which is 
solved 
analytically in first iteration of the quasi-linearization method (QLM). 
The zeroth iteration is based on general features of 
the exact solution near the boundaries.
The approach is illustrated on the 
Yukawa  potential. The 
results enable
accurate analytical estimates of effects of parameter variations on 
physical systems.
\end{abstract}

\maketitle

\thispagestyle{fancy}


We find accurate analytic presentation 
of wave functions and energies for an arbitrary physical 
potential $U(r)$. We use the quasilinearization method (QLM) 
suggested recently for solving the Schr\"{o}dinger equation
after 
conversion to Riccati 
form
\cite{A0,A2}. In QLM
 the nonlinear terms of the differential equation are 
approximated 
by a sequence of linear
expressions. The QLM is iterative but not perturbative and 
gives
stable solutions to nonlinear problems without depending on 
the
existence of a smallness parameter. 

Substitution of expression $y(r)=\frac{\chi'(r)}{\chi(r)}$  converts the radial 
Schr\"{o}dinger equation $[-\frac{1}{2m}\frac{d^2}{dr^2}+U(r)]\chi(r)=E\chi(r)$
into the nonlinear Riccati equation $y'(r)+y^2(r)=k^2(r), k^2(r)=2m[U(r)-E].$

The  corresponding QLM equation \cite{A0,A2} is
$y'_{n+1}(r)+2y_{n+1}y_{n}(r)=y_n^2(r)+k_n^2(r)$,
where $k_n^2$ is obtained from $k^2(r)$ by replacing there $E$  by 
energy of n-th iteration
$E_n=\frac{\int_0^\infty 
\chi^*_n(r)H(r)\chi_n(r)dr}{\int_0^\infty 
\chi_n^*(r)\chi_n(r)dr}, \chi_n(r)=C \exp\left(\lambda\int^r y_n(r') dr'\right)$.
Since the QLM iterations have 
very fast quadratic convergence \cite{A0,A2}, one can expect that 
even the first iteration which is given by an analytic expression \cite{A1}
\begin{widetext}
\begin{equation}\label{27}
y_{1}(r)=\int_{0}^{r}e^{2\int_{r}^{s}y_{0}(t)dt}
\left[y_{0}^{2}(s)+k^{2}(s)\right]ds
=\frac{1}{\chi^{2}_{0}(r)}\int_{0}^{r}\chi^{2}_{0}(s)
\left[y_{0}^{2}(s)+k_{0}^{2}(s)\right]ds.
\end{equation}
\end{widetext}
will be  accurate if  
the zeroth 
iteration  based on general features of solutions near the 
boundaries is  chosen.

For illustration 
we find the wave functions and binding 
energies of Yukawa potential
analytically in the first QLM iteration.
To estimate the precision of our analytic solution we solve
 the Schr\"{o}dinger equation 
 numerically as well.

The Yukawa potential $U(r)=-g\frac{e^{-\lambda r}}{r},~g>0 $ 
was suggested in the early days of quantum
mechanics for  description of nucleon interactions. 
During last decades the Yukawa potential  
have been used in atomic physics 
applications,
such as the screening of nucleon electromagnetic field by
electron cloud, or atoms under external pressure and in connection to  
 quark
interactions with parameters 
$\lambda$ 
and $g$ depending on the temperature of the quark-gluon plasma.

Let us try to guess the simplest form of the wave function. 
The large distance behavior is $\psi(r) \sim  e^{-\eta r}$
with $\eta=\sqrt{-2mE}$,
while the small distance behavior is determined by the Kato condition \cite{K} 
$\psi(0)=-\frac{\psi'(0)}{\mu}$ with $\mu=mg$. 
Noting also that the radial wave function should have a 
nonzero value at the origin, we come to the following initial guess function
$\psi_0(r) \sim \frac{e^{-\eta r}-e^{-ar}}{r}$
with $a$ chosen to satisfy the Kato condition which leads to 
$a=2\mu-\eta$. 
Thus we find for the initial guess  
$\chi_0(r)=r \psi_0(r)=N\left[ {e^{-\eta r}-e^{-(2\mu-\eta)r}}\right]$,
and therefore
 $y_{0}(r)=-\mu+\left(\mu-\eta\right) \coth 
\left[\left(\mu-\eta \right)r \right]$ 
where 
$N=\frac{\sqrt{\mu\eta\left(2\mu-\eta \right) }}{\mu-\eta }$
is the normalization factor.

Inserting 
this into the equation for  $E_0$ 
and 
using a straightforward integration, one obtains 
for the zeroth order ground state energy


\begin{widetext}
\begin{equation}\label{75a}
E_0=\frac{\mu\eta(2\mu-\eta)}{m}
\left[ \frac{1}{2\mu}+\frac{\mu}{\left(\mu-\eta \right)^2 }
\ln\frac{\left(4\mu-2\eta+\lambda \right)\left(2\eta+\lambda \right)  }
{\left(2\mu+\lambda \right)^2}
\right].
\end{equation}
\end{widetext}

Since $E_{0}(\eta)=-\frac{\eta^2}{2m}$ this is a 
 transcendental equation for 
 the parameter $\eta$.

The first iteration of the logarithmic derivative is given by 
Eq. (\ref{27}). Its explicit form is given by 
$y_1(r)=y_0(r)+\frac{\Phi(r)}{\chi_0^{2}(r)}$ where \cite{A1}
\begin{widetext}
\begin{eqnarray}\label{80}
\Phi(r)=2\mu \left\lbrace 
\left(\frac{\eta}{\mu}-1 \right)e^{-2\mu r}+
\frac{\mu-\eta}{2\mu-\eta}e^{-2r(2\mu-\eta)}+
\right.
~~~~~~~~~~~~~~~~~~~~~~~~\nonumber\\
\left.
2Ei\left[-r(2\mu+\lambda)\right]-Ei\left[-r(2\eta+
\lambda)\right]-Ei\left[-r(4\mu-2\eta+\lambda\right].
\right\rbrace 
\end{eqnarray}
\end{widetext}
Here $Ei(z)=-\int_{-z}^{\infty}\frac{e^{-t}}{t}dt$ is 
exponential integral function.
Inserting 
$y_{1}(r)$ 
into the expression for  $E_1$, one can calculate the first iteration
energy.

Numerical results for the binding energies are given in Table 1 
while the typical modified ground state wave function $\chi(r)$ 
is displayed on Fig. 1. We 
are using atomic 
system of units $m=g=1$. The
dimensionless parameter $\lambda$ is expressed in units of 
the inverse
Bohr radius while
the energy $E$ is expressed in Hartree.


Comparison of our approximate analytic expressions for 
binding energies and wave functions with the exact numerical 
solutions demonstrates their high accuracy in the wide range 
of physical parameters. The accuracy ranging between 
$10^{-4}$ and $10^{-8}$ for the energies and, 
correspondingly, $10^{-2}$ and $10^{-4}$ for the wave 
functions is reached. Similarly, the accurate analytic presentation of 
the solution of the Schr\"{o}dinger Equation could be 
obtained for an arbitrary potential.

\begin{table}
\caption{The energy values calculated by direct numerical solution of the wave 
equation ($E_D$), and in the zeroth and first iterations of QLM ($E_0$) and ($E_1$).}
\begin{center}
\begin{tabular}{|c|c|c|c|c|}
\hline
$\lambda $ & $-E_0$ & $-E_1$ & $-E_D$  \\
\hline
$0.2$ & 0.32679 & 0.32680851 & 0.32680851 \\
$0.5$ & 0.14795 & 0.1481170  & 0.1481170\\
 $0.8$ & 0.04445 & 0.0447042 & 0.0447043 \\
\hline
\end{tabular}
\end{center}
\vspace{-2mm} 
\end{table}

\begin{figure}
\begin{center}
\includegraphics[width=87mm]{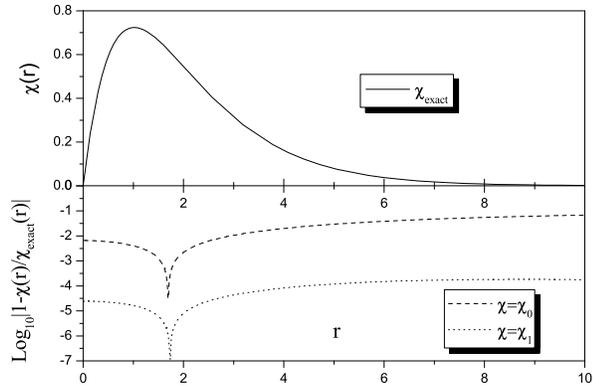}
\end{center}
\caption{Ground state wave functions for the Yukawa potential 
with parameter $\lambda=0.2$.
The exact modified wave function (solid line) is depicted at 
the upper part of the graph.
Relative logarithmic deviations from the exact values for the 
QLM wave functions of the zeroth (dash line) and the first 
order (dot line) are shown in the lower part.}
\label{fig2}
\end{figure}

\bigskip 

\begin{thebibliography}{9}   


\bibitem{A0} V.~B.\ Mandelzweig, J.\ Math.\ Phys.\ Com. 
\textbf {40}, 6266 (1999).
\bibitem{A2} V.~B.\ Mandelzweig, and F.\ Tabakin, Comp.\ 
Phys.\ Com. \textbf {141}, 268 (2001).
\bibitem{A1} E. Z.\ Liverts, E. G.\ Drukarev and V.~B.\ Mandelzweig, Annals
of Physics,
in press, 2007.
\bibitem{K} T. Kato, Com. Pure Appl. Math. {\bf 10}, 151 
(1957).


\end{thebibliography}

\end{document}